\documentclass{osa-article}

\journal{osajournal}

\articletype{Research Article}

\usepackage{amsmath,amsfonts,amssymb}
\usepackage{graphicx}
\usepackage{tablefootnote}
\usepackage[final]{changes}

\setcopyright{}

\begin{document}

\title{Aerogel scattering filters for cosmic microwave background observations}

\author{Thomas Essinger-Hileman,\authormark{1,*} Charles L. Bennett,\authormark{2} Lance Corbett,\authormark{2} Haiquan Guo,\authormark{3} Kyle Helson,\authormark{1} Tobias Marriage,\authormark{2} Mary Ann B. Meador,\authormark{4}, Karwan Rostem,\authormark{1} Edward J. Wollack\authormark{1}}

\address{\authormark{1}NASA Goddard Space Flight Center, Greenbelt, MD 20771 USA\\
\authormark{2}Department of Physics and Astronomy, Johns Hopkins University, Baltimore, MD 21218, USA\\
\authormark{3}Ohio Aerospace Institute, Cleveland, Ohio, 44142, USA\\
\authormark{4}NASA Glenn Research Center, 21000 Brookpark Road, Cleveland, Ohio 44135, United States}

\email{\authormark{*}thomas.m.essinger-hileman@nasa.gov} 

\begin{abstract}
We present the design and performance of broadband and tunable infrared-blocking filters for millimeter and sub-millimeter astronomy composed of small scattering particles embedded in an aerogel substrate. The ultra-low-density (typically $<$ 150 mg/cm$^3$) aerogel substrate provides an index of refraction as low as 1.05, removing the need for anti-reflection coatings and allowing for broadband operation from DC to above 1 THz. The size distribution of the scattering particles can be tuned to provide a variable cutoff frequency. Aerogel filters with embedded high-resistivity silicon powder are being produced at 40-cm diameter to enable large-aperture cryogenic receivers for cosmic microwave background polarimeters, which require large arrays of sub-Kelvin detectors in their search for the signature of an inflationary gravitational-wave background.
\end{abstract}

\section{Introduction}
\label{sec:intro}  

Millimeter-wave cryogenic receivers with bolometric detectors require rejection of infrared (IR) radiation to reduce thermal loads on the cold stages of the cryostat. As telescope apertures increase in diameter to accommodate larger focal planes, the requirements for infrared filtering become more stringent, while the fabrication of filters becomes correspondingly more difficult. This is a particular problem for current-generation cosmic microwave background receivers, which have receiver apertures approaching 1 m in diameter~\cite{2014SPIE.9153E..1IE, 2014SPIE.9153E..1PB, 2016SPIE.9914E..0SG, 2016JLTP..184..805S, 2016ApJS..227...21T}. For such large apertures, IR rejection at 1 part in $10^6$ is required to enable operation of the focal plane array at sub-Kelvin temperatures. 

A variety of approaches have been developed to meet this challenge. Reflective metal-mesh filters, composed of patterned metal films on a thin dielectric substrate, have been widely used~\cite{1967InfPh...7...37U, 1967InfPh...7...65U, 1985ApOpt..24..217W, 2006SPIE.6275E..26A, 1981ApOpt..20.1355T}. Capacitive grids provide strong ($> 30 $ dB) IR rejection with minimal loss at millimeter wavelengths, though there are diminishing returns when using multiple reflective filters as reflections from subsequent filters need to transmit through filters further up in the stack to be rejected out of the receiver. Absorptive filters have also been extensively used, in which a material with low loss at millimeter wavelengths but strong absorption or reststrahlen reflection in the IR absorbs power and conducts it to higher-temperature stages of a receiver~\cite{Halpern1986, Bock:95, Munson:17}.  Materials used for this purpose include polytetrafluoroethylene (PTFE), nylon, and alumina. Absorbing filters require anti-reflection (AR) coatings and lose effectiveness as their diameter increases, because the centers of the filters tend to heat and re-radiate further down the optical chain. Low-refractive-index foam materials have been used, which scatter and absorb IR power due to their pore size, and then radiate some of that absorbed power back out of the receiver~\cite{2013RScI...84k4502C}.

We here present the design, fabrication, and performance of IR-blocking filters composed of small scattering particles embedded in an aerogel substrate. They are made to diffusely scatter infrared radiation to a wide range of angles~\cite{DeVore:47, Plummer:36}. The ultra-low-density (90-200 mg/cm$^3$) aerogel substrate has a low index of refraction, removing the need for an AR coating and allowing for high transmission across an ultra-broad band from zero frequency to greater than 10 THz. The size of the scattering particles can be tuned to give variable cutoff frequencies. 
  \begin{figure} [t!]
  \begin{center}
  \includegraphics[width=0.6\textwidth, clip=true, trim=2.5in 2.6in 3in 1.5in]{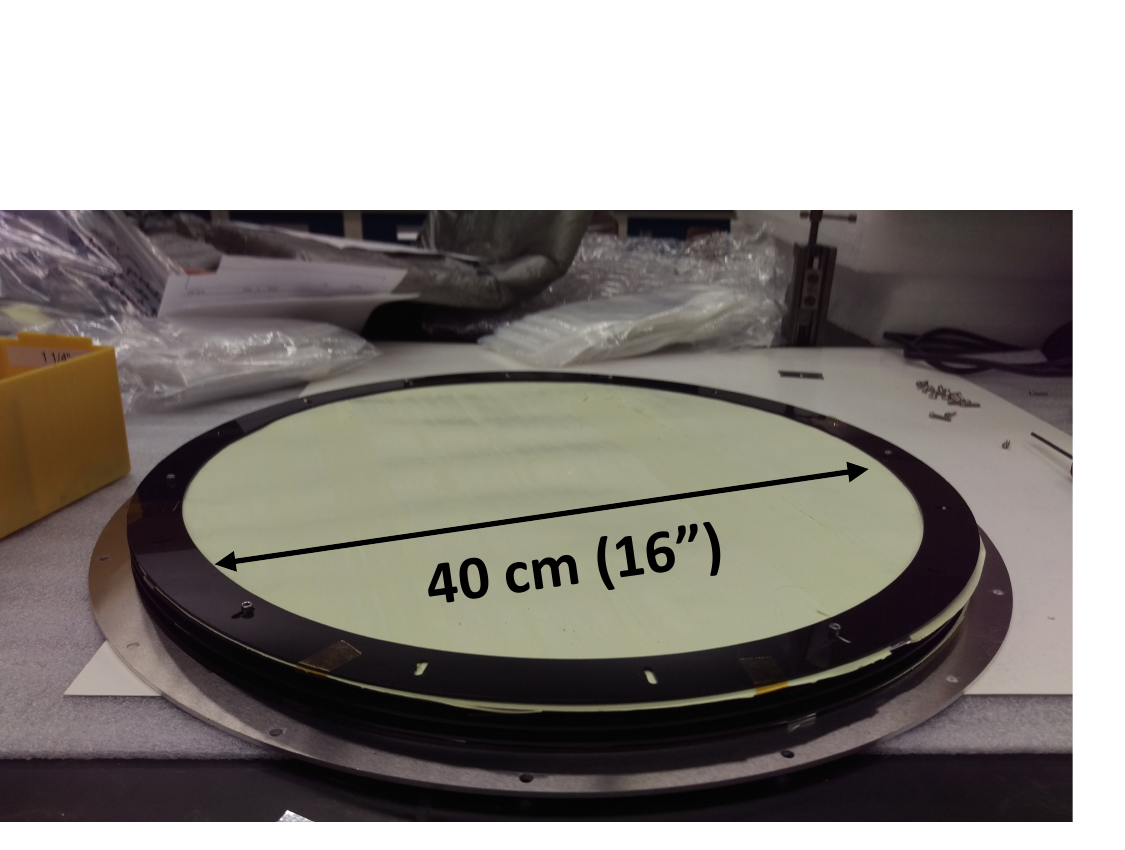}  
  \end{center}
  \caption[example] 
  { \label{fig:polyimide_photo} 
\added{Photograph of a 40-cm-diameter polyimide aerogel scattering filter mounted on a ring for installation in a receiver.}}
  \end{figure} 

\begin{table}[tp]
\begin{center}
\begin{tabular}{|c|c|c|c|c|c|c|c|} \hline
 & Dimensions & Aerogel & Aerogel & Si Part. & Si Part. & Model & Meas.  \\
 & & Material & Density & Sizes & Density & $n$ & $n$ \\ \hline 
 & mm & & mg/cm$^3$ & $\mu$m & mg/cm$^3$ & & \\ \hline 
 1 & 3 x 50 x 50 & Silica & 90 & None & N/A & 1.03 & 1.04 \\ \hline
 2 & 3 x 50 x 50 & Silica & 90 & 60-75 & 35 & 1.04 & 1.07 \\ \hline
 3 & 3 x 50 x 50 & Silica & 90 & 60-75 & 17 & 1.03 & 1.06 \\ \hline
 4 & 12.5 x 50 x 50 & Silica & 90 & 75-100 & 35 & 1.04 & ---\tablefootnote{High accuracy index of refraction data were not able to be obtained on this thicker sample.} \\ \hline
 5 & 0.3 x 420 x 2000  & Polyimide 1 & 212 & None & N/A & 1.09 & 1.10 \\ \hline
 6 & 0.5 x 420 x 2000  & Polyimide 1 & 178 & 50-75 & 3 & 1.07 & 1.08 \\ \hline
 7 & 0.3 x 420 x 2000  & Polyimide 1 & 204 & 75-100 & 5 & 1.09 & 1.13 \\ \hline
 8 & 1.0 x 25 x 25 & Polyimide 2 & 125 & None & N/A & 1.05 & 1.10 \\ \hline
 9 & 0.9 x 25 x 25 & Polyimide 2 & 125 & $< 30$ $\mu$m & 50 & 1.08 & 1.11 \\ \hline
 10 & 1.2 x 25 x 25 & Polyimide 2 & 125 & $< 30$ $\mu$m & 200 & 1.16 & 1.21 \\ \hline
\end{tabular}\vspace{10pt}
\caption{Summary of samples measured. The model assumes index of refraction of 1.7, 2.0, and 3.4 for polyimide, silica, and silicon, respectively. The silicon powder was produced from p-type, boron doped, float-zone silicon wafers with resistivity $> 10$ k$\Omega$-cm by pulverizing by hand in a mortar and pestle or using a ball milling machine and then sieving to produce powder with a given size distribution. Two polyimide aerogel formulations were investigated. See the text for a detailed description of the two polyimide formulations. The samples show measured indices of refraction above that predicted from our models (See Sec.~\ref{sec:maxwell_garnett}) given the bulk properties of silica, polyimide, and silicon alone. This may indicate a difference between the bulk properties and the properties of the materials in the gel matrix or the presence of other adsorbed materials, such as water, facilitated by the high pore surface area of the aerogels.  \label{tbl:samples}}
\end{center}
\end{table}%

Aerogel materials investigated for this work were a classic silica aerogel and flexible and mechanically robust polyimide aerogels made with two different polyimide oligomer backbones. Polyimide 1 was made from 9 \% w/w polymer concentration, using biphenyl-3,3$^{\prime}$,4,4$^{\prime}$-tetracarboxylic dianydride (BPDA) as dianhydride, the combination of 2,2$^{\prime}$-dimethylbenzidine (DMBZ) and 4, 4$^{\prime}$-oxydianiline (ODA) as diamine, and 3,5-triaminophenoxybenzene (TAB) as cross-linker.  Polyimide 2 was made from 10 \% w/w polymer concentration, using BPDA as dianhydride, the combination of 4,4$^{\prime}$-bis (4-aminophenoxy) biphenyl (BAPB) and DMBZ as diamine, and TAB as cross-linker. 

Silica aerogel samples 12 mm $\times$ 50 mm $\times$ 50 mm were successfully fabricated with densities of 90 mg/cm$^3$ and silicon loading of 17 and 35 mg/cm$^3$ with silicon particle size distributions of 60-75 $\mu$m and 75-100 $\mu$m. Polyimide aerogel 1 was produced in rolls 43 cm $\times$ 2 m in thicknesses of 0.3-0.5 mm with silicon loading of 3 and 5 mg/cm$^3$ with particle size distributions of 50-75~$\mu$m and 75-100~$\mu$m. Polyimide aerogel 2 was produced in smaller coupons 25 mm x 25 mm with thicknesses of 0.9-1.2 mm, silicon loading of 50 and 200 mg/cm$^3$, and particle size distribution of $< 30$ $\mu$m. Reference samples of unloaded aerogel were produced for both silica and polyimide aerogels. 

\section{Filter Fabrication and Sample Description}
\label{sec:fabrication}
Aerogels are made by supercritical drying of a gel, typically using carbon dioxide. Silicon powder is mixed into the gel before curing and supercritical drying. Silica aerogel samples were produced by Ocellus, Inc.\footnote{Ocellus, Inc., 450 Lindbergh Avenue, Livermore, CA 94551, information@ocellusinc.com} Silica aerogel samples were initially produced in two base aerogel densities of 50 and 90 mg/cm$^3$ with a loading of 75-100 $\mu$m silicon particles at a density of 35 mg/cm$^3$. The two initial samples were 50 mm x 50 mm x 12.5 mm in size. The 50 mg/cm$^3$ sample was found to be too fragile to reliably mount in sample holders for optical measurements, so all further development focused on 90 mg/cm$^3$ aerogel density. Three further samples were produced and measured with sizes of 50 mm x 50 mm x 3 mm, labeled Samples 1--3, as summarized in Table~\ref{tbl:samples}. A reference aerogel was produced with no silicon powder and samples with 17 mg/cm$^3$ and 35 mg/cm$^3$ silicon loading density with a particle size distribution of 60--75~$\mu$m were produced.  The initial thicker sample is labeled Sample 4. 

Polyimide aerogel samples were produced at NASA Glenn Research Center as described in previous publications~\cite{Meador2012a, Meador2012b, Guo2012}. Polyimide 1 oligomers were formulated in N-methylpyrolidone (NMP) by using 26 equivalents of dianhydride and 25 equivalent diamine.  Then they were cross-linked using triamine based TAB.  BPDA was the dianhydride.  50 mol\% DMBZ and 50 mol\% ODA were added together as diamine to make random polyimide oligomers.  The total polymer concentration was 9 \% w/w.  Silicon powders were added right after adding TAB.  Then Acetic anhydride and triethyleneamine were used to chemically imidize the polyamic acids at room temperature.  The polyimide aerogel rolls were produced through a roll-to-roll process, in which the sol was cast to a carrier plastic film in a uniform thickness, allowed to gel, and then dried using supercritical extraction of CO$_2$. 

Three rolls of Polyimide 1 were produced with a nominal density of 200 mg/cm$^3$ in sizes approximately 42 cm x 2 m and thicknesses in the range 0.3--0.5 mm. 
Sample 5 is a reference sample with no silicon powder, Sample 6 was loaded with 50-75~$\mu$m silicon powder, and Sample 7 was loaded with 75--100~$\mu$m silicon powder. The addition of silicon powder may decrease shrinkage of the aerogel during supercritical drying, but further investigation is needed to establish this. 

Three further samples were produced at small scale, approximately 1 mm x 25 mm x 25 mm. These samples aimed to increase IR rejection and improve mm-wave transmission using a polyimide aerogel formulation, Polyimide 2, with lower base aerogel density $\sim 125$ mg/cm$^3$, smaller silicon powder particle size $< 30$ $\mu$m, and higher silicon loading density. Polyimide 2 oligomers were made through the same procedure as described for polyimide 1 except instead of using 50 mol\% ODA and forming random polyimide oligomers, 50 mol\% BAPB was reacted with BPDA first then 50 mol\% DMBZ was added later to form alternated polyimide oligomers. Sample 8 was the base polyimide aerogel, while Samples 9 and 10 additionally contained 50 mg/cm$^3$ and 200 mg/cm$^3$, respectively, of silicon powder.

\begin{figure} [t!]
   \begin{center}
   \begin{tabular}{cc} 
   \includegraphics[width=0.46\textwidth, clip=true, trim= 0in 0.1in 0in 0in]{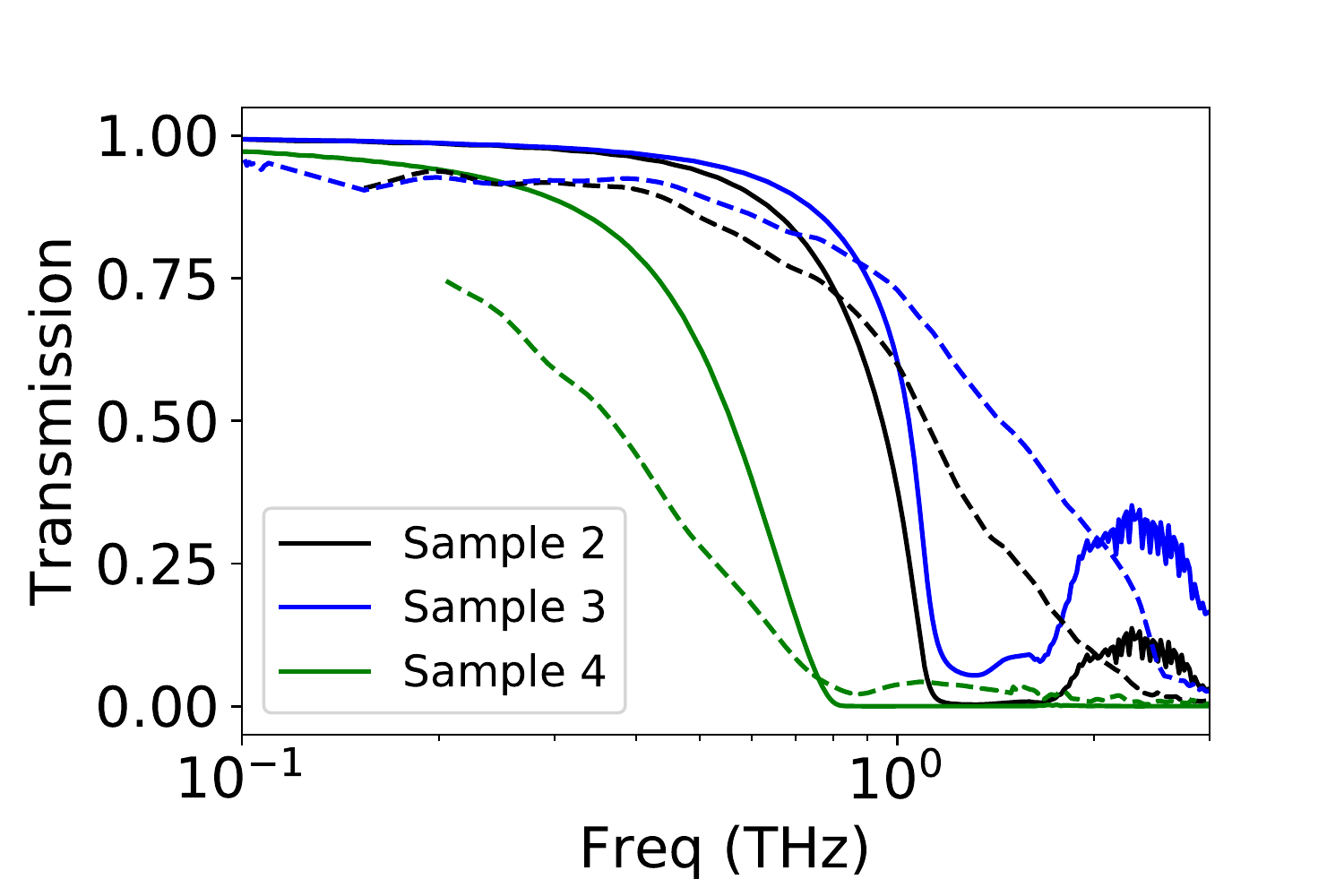} &
   \includegraphics[width=0.46\textwidth]{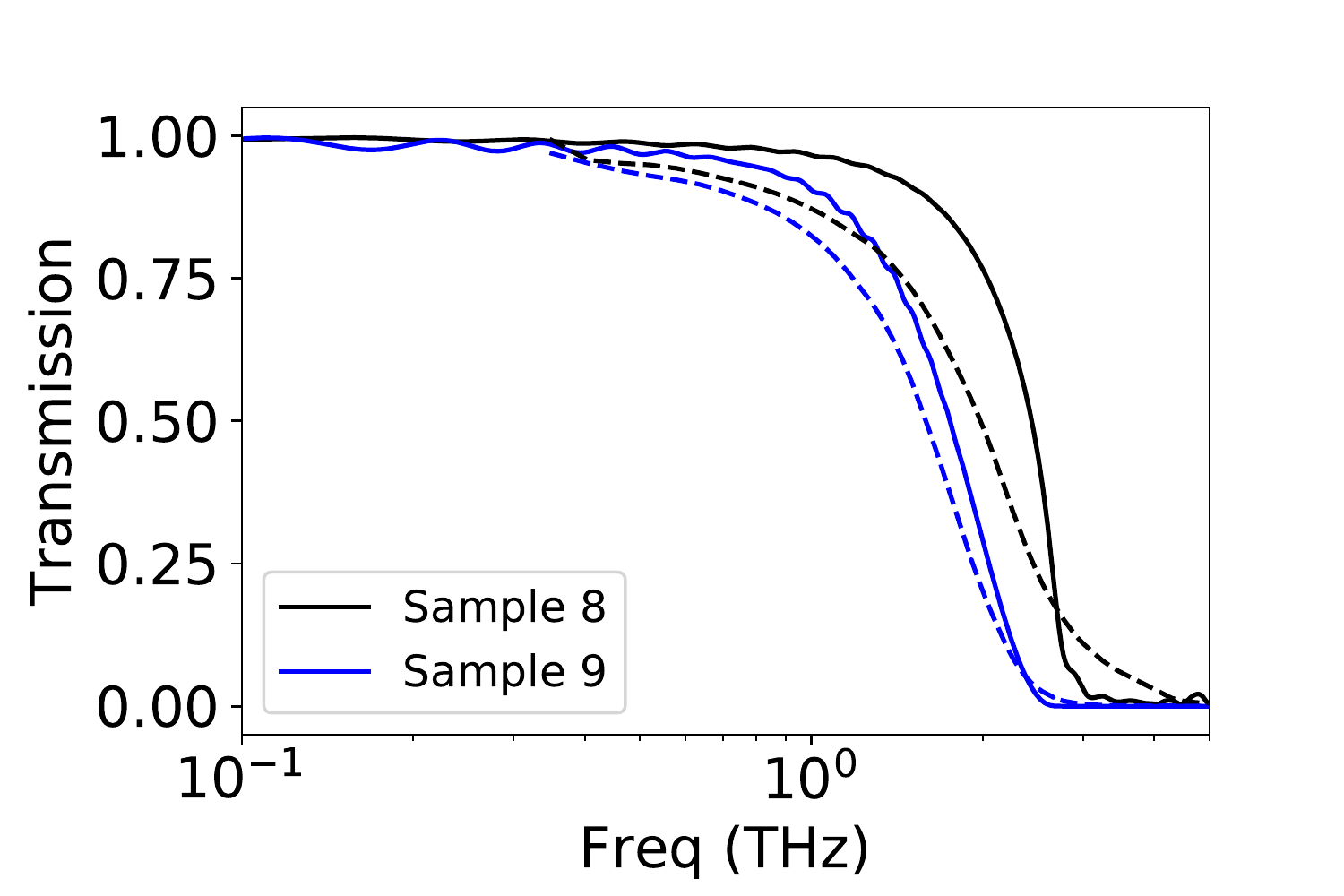}   
   \end{tabular}
   \end{center}
   \caption[example] 
   { \label{fig:mm_trans_model} 
\added{Examples of modeled transmission curves (solid) versus measured transmission curves (dashed) for aerogel scattering filter samples. Sample numbers follow those in Table~\ref{tbl:samples}.  All models are for the full integrated model of Section~\ref{sec:integrated_model} and assume a loss tangent of $\tan\delta = 10^{-3}$. We emphasize that the models are not fit to the data. They assume the parameters of Table~\ref{tbl:samples} with no free parameters. \textit{Left:} Results for silica aerogel samples. All samples have base aerogel density of 90~mg/cm$^3$. Sample 2 (black curves) is 3~mm thick and is loaded with 35~mg/cm$^3$ of 60--75~$\mu$m silicon particles; Sample 3 (blue curves) is 3~mm thick with 17~mg/cm$^3$ of 60--75~$\mu$m silicon particles; Sample 4 (green curves) is 12.5~mm thick with 35~mg/cm$^3$ of 75--100~$\mu$m silicon particles. Note that behavior of the model above cutoff, with an observed harmonic leak, is not seen in measured data, likely due to the non-spherical shape of the particles. \textit{Right:} Model (solid) versus measurement (dashed) for polyimide aerogel scattering filter Samples 8 (black curves) and 9 (blue curves), which are approximately 1~mm thick with a base aerogel density of 125~mg/cm$^3$, loaded with silicon particles with sizes sieved to be $< 30$~$\mu$m in size. The samples differ only in density of particle loading, which is 50~mg/cm$^3$ for Sample 8 and 200~mg/cm$^3$ for Sample 9.}}
\end{figure} 

\section{Optical Design}

\subsection{Low-frequency performance}\label{sec:maxwell_garnett}
The filter response is modeled using a combination of effective-dielectric and Mie scattering theory. \added{Example model outputs compared with measurements are shown in Fig.~\ref{fig:mm_trans_model}. The model is primarily used to choose formulations to achieve a given cutoff frequency. Further discussion of the agreement between model and measurement is reserved for Section~\ref{sec:model_meas}.} For wavelengths that are large compared to the scattering particle size and separation, the aerogel plus scatterers will behave as a composite material of air, the aerogel matrix material, and silicon. The aerogel without scatterers will have an approximate effective dielectric constant given by Maxwell-Garnett theory~\cite{MaxwellGarnett1904, Niklasson1981, Sihvola2008} of:

\begin{equation}
\epsilon_{a} = \frac{\epsilon_{m} + 2 + 2 f_a \left( \epsilon_{m} - 1 \right)}{\epsilon_{m} + 2 - f_a \left( \epsilon_{m} - 1 \right)},
\label{eqn:maxwell_garnett1}
\end{equation}

\noindent where $\epsilon_{a}$ and $\epsilon_{m}$ are the dielectric constants of the aerogel and the aerogel matrix material, respectively, and $f_a$ is the volume filling fraction of material in the aerogel. The nanoporous structure of typical aerogels allows them to behave like effective dielectrics well into the infrared and on length scales associated with the scattering particles used in this work ($>$ 1 $\mu$m).

The inclusion of a small filling fraction of scattering particles increases the dielectric constant. For the small filling fractions ($\lesssim 10$\%) used in the scattering filters, it is valid to calculate an effective dielectric constant for the base aerogel and then use Maxwell-Garnett theory again to calculate a new effective dielectric constant 

\begin{equation}
\bar{\epsilon} = \epsilon_{a} \frac{\epsilon_{s} + 2 \epsilon_{a} + 2 f_s \left( \epsilon_{s} - \epsilon_{a} \right)}{\epsilon_{s} + 2 \epsilon_{a} - f_s \left( \epsilon_{s} - \epsilon_{a} \right)},
\label{eqn:maxwell_garnett2}
\end{equation}

\noindent where $\epsilon_a$ is the aerogel dielectric constant calculated from Eq.~\ref{eqn:maxwell_garnett1}, $\epsilon_s$ is that of the scattering particles, and $f_s$ is the volume filling fraction of the scattering particles in the medium. We note that at frequencies where the materials involved are not overly lossy, the Maxwell-Garnett formulas, Eq~\ref{eqn:maxwell_garnett1} and Eq~\ref{eqn:maxwell_garnett2}, can be used to estimate loss in the composite material by inserting complex dielectric constants for the materials.

   \begin{figure} [t!]
   \begin{center}
   \begin{tabular}{cc} 
   \includegraphics[width=0.5\textwidth, clip=true, trim= 0in 0.1in 0in 0in]{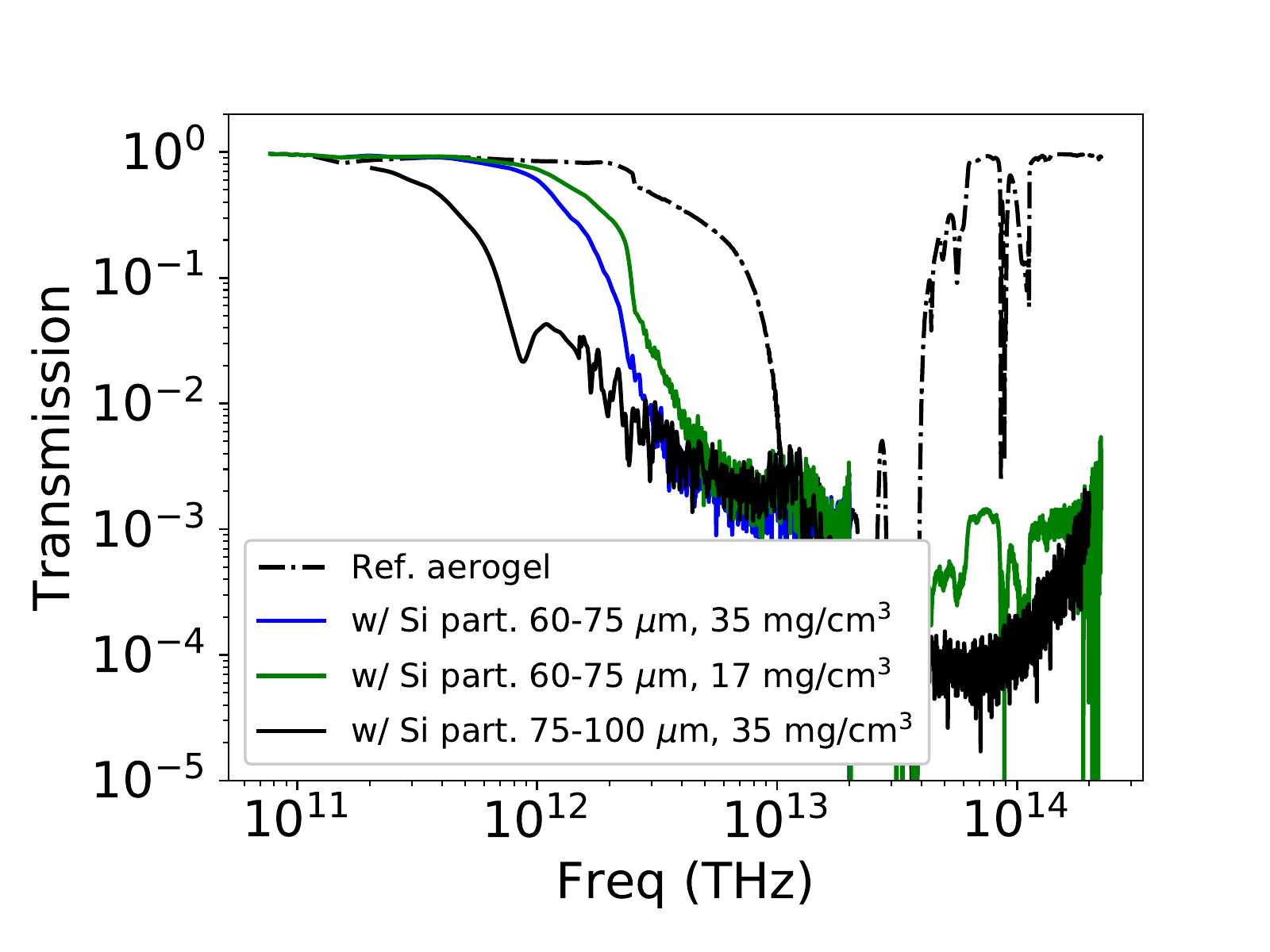} &
   \includegraphics[width=0.4\textwidth]{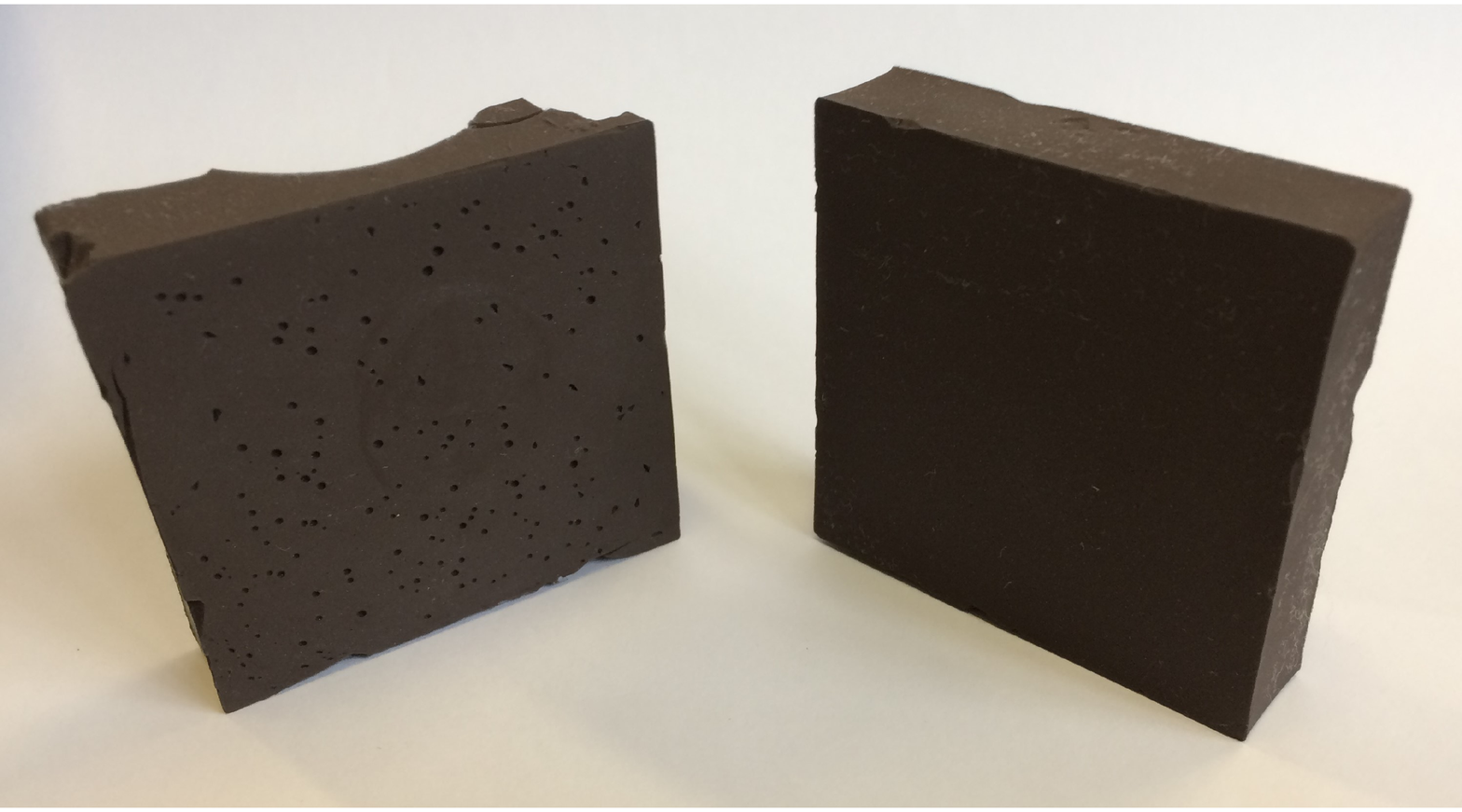}   
   \end{tabular}
   \end{center}
   \caption[example] 
   { \label{fig:silica_trans_photos} 
\textit{Left:} Transmission spectra of 90 mg/cm$^3$ silica aerogel with no silicon powder (dot-dash black), with 17 mg/cm$^3$ (blue) and 35 mg/cm$^3$ (green) of 60-75 $\mu$m silicon particles, and with 35 mg/cm$^3$ of 75-100~$\mu$m silicon particles (solid black). Samples were approximately 3 mm thick. Measurements below 1.5 THz were made with a time-domain THz spectrometer, while measurements at higher frequencies were made with a Bruker Fourier transform spectrometer. \added{Note that the silica reference sample exhibits strong absorption bands $\sim$ 5--150~THz, but high transmission above 150~THz. Scattering from the pores of the silica aerogel itself will only become important in the visible, given the $\sim 20$~nm pore size of the silica aerogel.} \textit{Right:}~Photographs of two silicon-loaded silica aerogel samples of different silica density. The left sample has density of 50 mg/cm$^3$ and the right sample 90 mg/cm$^3$.}
   \end{figure} 

The reflection at normal incidence for such a film with a thickness $t$ is calculated in standard texts~\cite{1999prop.book.....B}:

\begin{equation}
R = \frac{2 r^2 \left( 1 + \cos 2 \beta \right)}{1 + r^4 + 2 r^2 \cos 2 \beta},
\label{eqn:dielectric_sheet}
\end{equation}

\noindent where $r = (\bar{n}-1)/(\bar{n}+1)$ for $\bar{n} = \sqrt{\bar{\epsilon}} \simeq \sqrt{  \operatorname{Re}(\bar{\epsilon})}$, where the final equality is approximately true when $\operatorname{Im}(\bar{\epsilon}) \ll \operatorname{Re}(\bar{\epsilon})$, a valid approximation for our materials. $\beta = 2 \pi \nu \bar{n} t / c$ is the phase delay in the material at frequency $\nu$. For small $r$, the total reflectance exhibits fringes between $0$ and $4 r^2$. The aerogels in this study have indices of refraction, $n$, in the range 1.04-1.13. For a material in the middle of this range with $\bar{n} = 1.10$, the maximum reflectance is 0.9\% and the average reflectance across all frequencies is $<0.5$\%. Judicious choice of filter thickness for a given observing band can reduce reflections from the filter to levels of 0.1\%.

The samples show measured indices of refraction above that predicted from Maxwell-Garnett theory given the bulk properties of silica, polyimide, and silicon alone. This may indicate a difference between the bulk properties and the properties of the materials in the gel matrix or the presence of other adsorbed materials, such as water, facilitated by the high pore surface area of the aerogels. \added{Subsequent modeling can take this increased index of refraction into account for understanding its projected impact on future samples made via the same formulation. Further testing is planned to investigate the effect of water adsorption on filter properties in future work.}

\subsection{Scattering at high frequency}\label{sec:mie_scattering}
Optical scattering theory has been developed extensively in the literature~\cite{1957lssp.book.....V, 1983asls.book.....B}. The scattering of light from spherical particles can be solved analytically~\cite{1908AnP...330..377M}. The scattered intensity can be expanded as a sum of Bessel functions and depends on the ratio between the size of the spherical particle and the wavelength of light, as well as the scattered direction and indices of refraction of the medium and the scatterers. For a medium containing many small scatterers with wide separation, transmission through the medium can be shown to depend solely on the forward-scattered amplitude, $Q_{\text{sca}}$, from a single particle. See Appendix~\ref{sec:scattering_theory} for further details.

The specific intensity of light passing through a medium of scatterers is attenuated as 

\begin{equation}
I (z) = I_{0} \exp \left( - N \pi a^{2} Q_{\text{sca}} z \right) = I_{0} \exp \left( - \gamma z \right),
\end{equation}

\noindent where N is the volume density of scatterers, $a$ is the particle radius, and $z$ is the distance traveled through the medium. The attenuation coefficient has been defined as $\gamma=N \pi a^{2} Q_{\text{sca}}$.

The attenuation through a scattering medium with a distribution of sizes can be straightforwardly handled by integrating over the particle size. If the number of scatterers per unit volume is a function of particle radius, $N(a)$, then the attenuation coefficient \added{at a given wavelength} is 
\vspace{-10pt}

\begin{equation}
\gamma = \int_{0}^{\infty} N(a) \pi a^{2} Q_{\text{sca}} (2 \pi a / \lambda) da .
\end{equation}

   \begin{figure} [t!]
   \begin{center}
   \begin{tabular}{cc} 
   \includegraphics[width=0.50\textwidth]{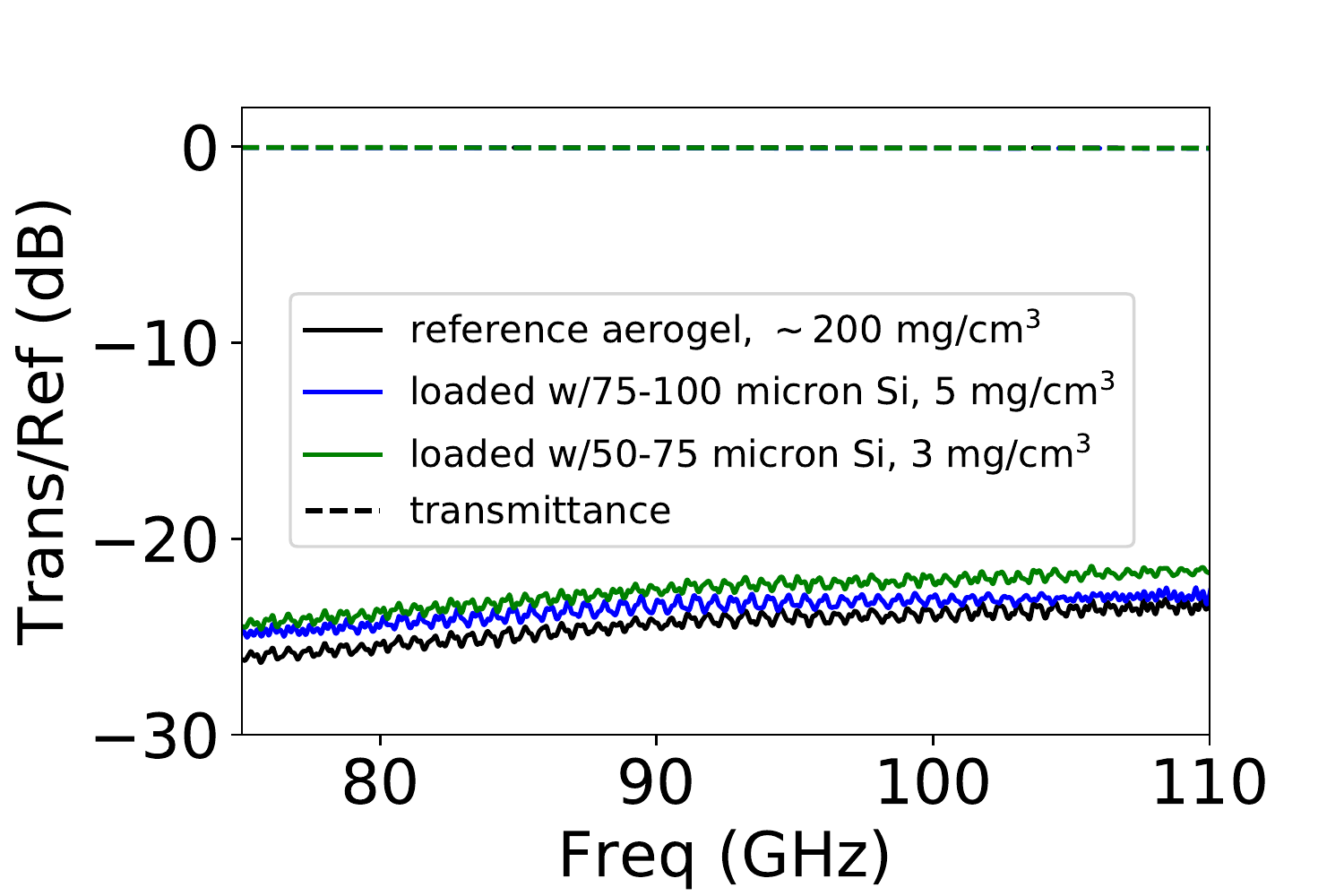} & 
   \includegraphics[width=0.42\textwidth, clip=true, trim=1in 0.5in 0.9in 1in]{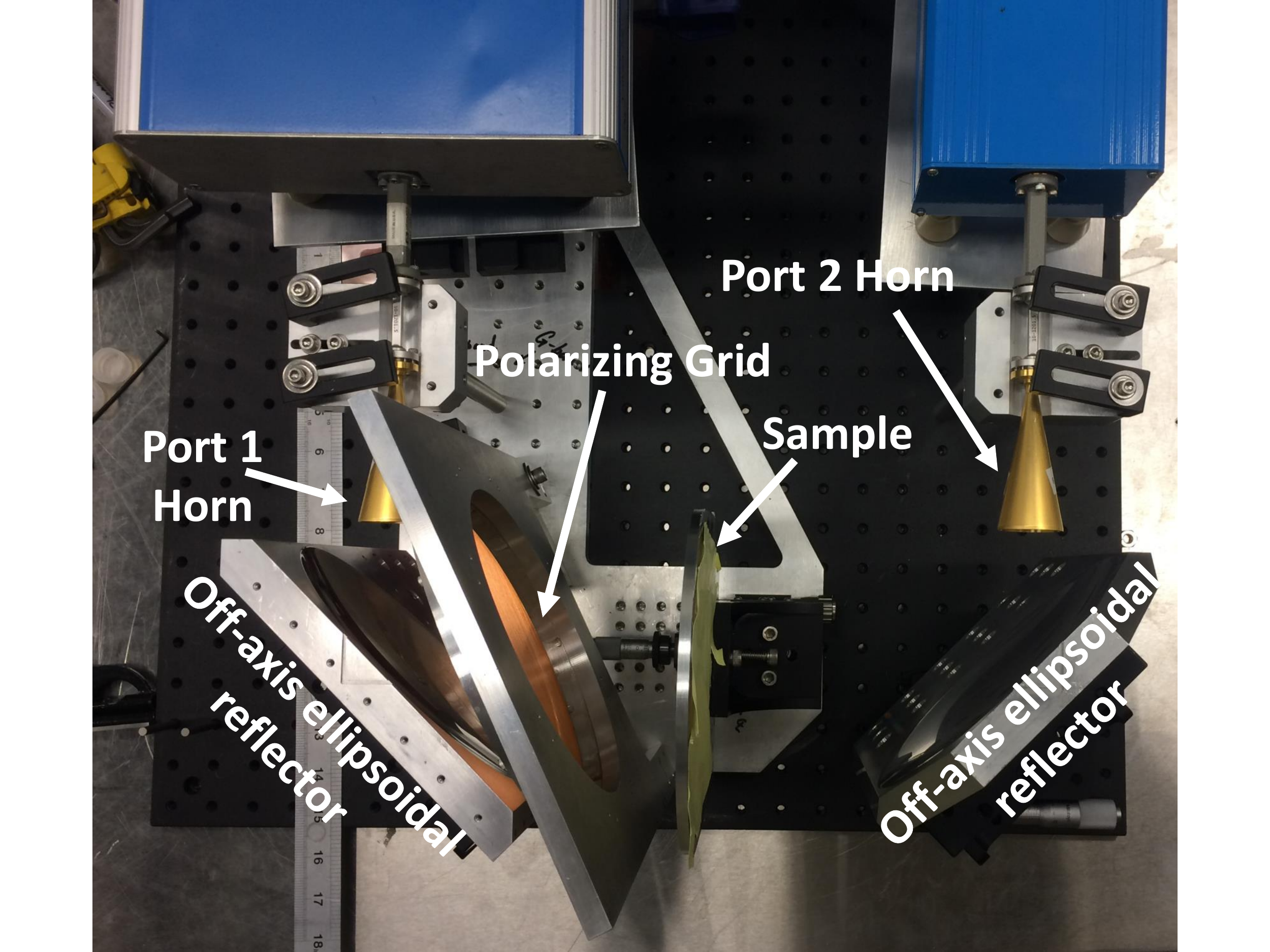}   
   \end{tabular}
   \end{center}
   \caption[example] 
   { \label{fig:polyimide_vna_meas} 
\textit{Left:} Vector network analyzer measurements of the reflectance (solid lines) and transmittance (dashed lines) of polyimide aerogel samples with no silicon powder (black), 3 mg/cm$^3$ of 50-75~$\mu$m silicon powder (green), and 5 mg/cm$^3$ of 75-100~$\mu$m silicon powder (blue). \textit{Right:} Photograph of the setup used to make the measurements. Scalar feedhorns both transmit and receive power through the system. Off-axis ellipsoidal reflectors provide a beam waist at approximately the location of the sample and a polarizing wiregrid reduces mode coupling in the setup. The sample is mounted on a linear translation stage. }
   \end{figure} 

Using this model, we are able to estimate the frequency cutoff of our filters in the limits that:

\begin{enumerate}
\item The particles are roughly spherical. This is not strictly true, but is a reasonable first approximation for randomly-oriented nearly-spherical particles and is particularly relevant for long wavelengths.
\item Scattering is not strong enough for multiple scattering events to play a significant role. 
\end{enumerate}

\subsection{Integrated Model}
\label{sec:integrated_model}
The complex dielectric constant, $\bar{\epsilon}$, estimated from Maxwell-Garnett theory in Sec.~\ref{sec:maxwell_garnett} can be combined with the additional attenuation, $\gamma$, estimated from Mie scattering theory in Sec.~\ref{sec:mie_scattering}. The attenuation due to scattering contributes to the complex part of the dielectric constant to give a total effective dielectric constant of

\begin{equation}
\epsilon_{eff} = \operatorname{Re}(\bar{\epsilon}) + \imath \left( \operatorname{Im}(\bar{\epsilon}) + \frac{c \gamma }{ 2 \pi \bar{n} \nu }\right) ,
\end{equation}

\noindent where the first term in parentheses arises from the intrinsic loss of the materials in the filter and the second term is extrinsic loss from scattering due to the geometry of the scattering particles.

This dielectric constant can then be incorporated into a transfer-matrix code for calculation of the expected transmission spectrum of a given filter. An example of model outputs for silica aerogel filters 5 mm thick with silicon particles embedded at a density of 35 mg/cm$^3$ of three different sizes (25~$\mu$m, 50~$\mu$m,  and 75~$\mu$m) is shown in Fig.~\ref{fig:mm_trans_model}. We note that because our particles are not spherical and multiple scattering plays a role at high frequencies, the model does not reproduce filter response in great detail. Rather the model is used to guide filter design to choose appropriate particle size and loading volume percentage to achieve a desired cutoff frequency. 

\section{\replaced{Testing}{Measurements}}

Transmission and reflection data were taken of samples to evaluate their performance. The silica aerogel samples were measured in transmission using a time-domain THz spectroscopy system~\cite{2012OExpr..2012303M} for frequencies 0.15--1.5 THz and a Bruker FTS for transmission at frequencies of 1.5--200 THz. Polyimde aerogel samples were measured using a Bruker FTS for infrared transmission across frequencies 1--18 THz and a quasi-optical vector network analyzer (VNA) setup for reflection at W band (75-110 GHz). VNA reflectance measurements were taken multiple ($\sim 5$) times with sample positions translated by approximately $\lambda/5$ each time, following the method outlined in Ref.~\citenum{2011RScI...82h6101E}. \added{Different measurement setups were used for silica and polyimide aerogel samples, because the measurements were made approximately one year apart, during which time the first author had changed institutions. Care was taken to properly calibrate all measurements to ensure comparable results between setups.}

Silica aerogel filters were mounted in holders with 8~mm diameter apertures after being cut to size with a diamond wire saw. Stycast 2850 epoxy was mixed and allowed to get tacky before application, as it was found that otherwise the epoxy wicked into the aerogel samples and caused fracturing. Polyimide aerogel samples were mounted on sample holders with aperture diameters of 125~mm for VNA measurements and 12.5~mm for FTS measurements using double-sided pressure-sensitive adhesive tape. 

\begin{figure} [t!]
  \begin{center}
  \includegraphics[width=0.50\textwidth, clip=true, trim=0in 0in 0in 0in ]{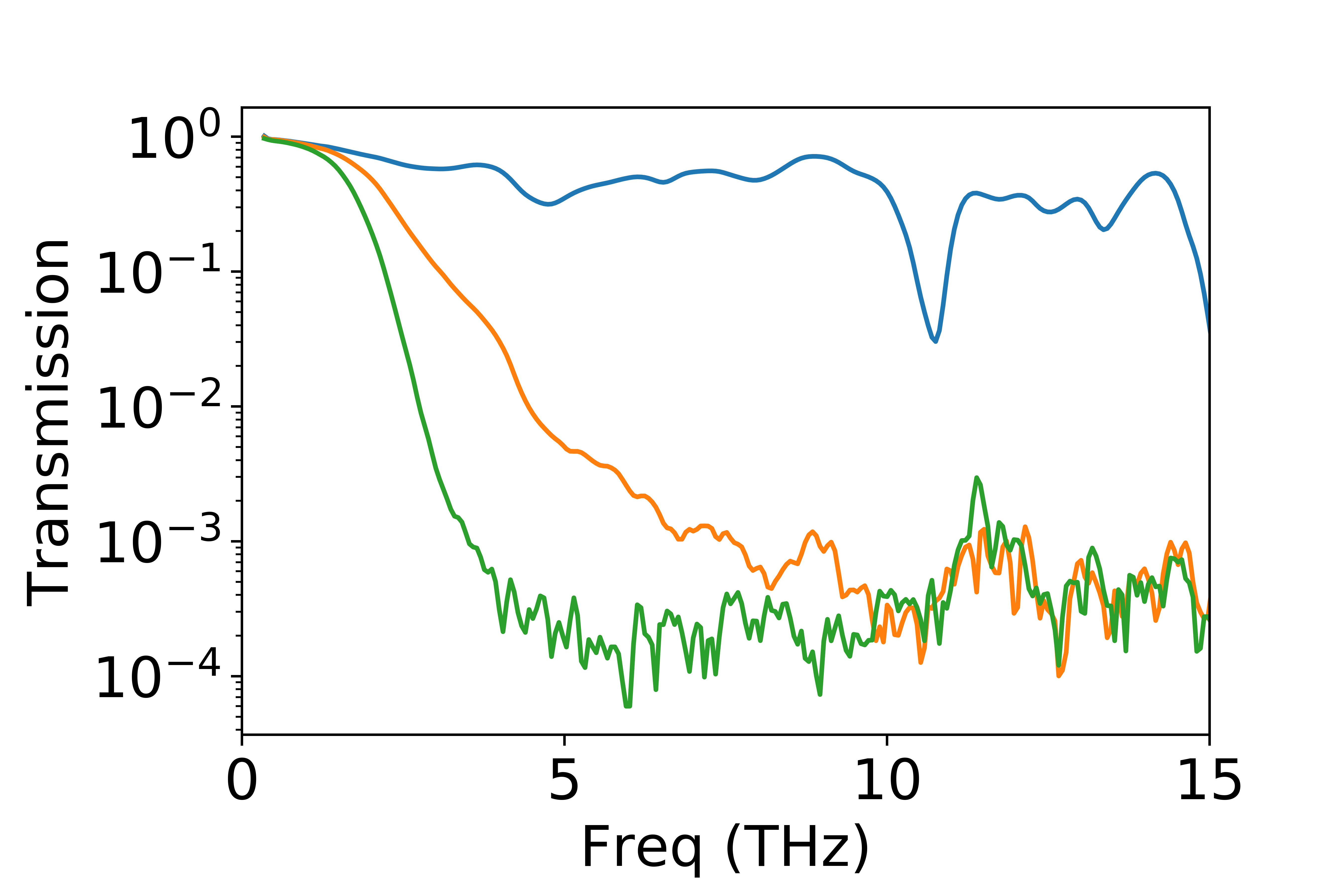}
  \end{center}
  \caption[example] 
  { \label{fig:polyimide_trans} 
\added{Measurements of IR transmission of polyimide aerogel samples with a Bruker FTS. Measurements were taken with a 12.5~mm diameter aperture for Samples 8, 9, and 10 with no silicon powder (blue), 50 mg/cm$^3$ silicon powder with size distribution $<30$ $\mu$m (orange), and 200 mg/cm$^3$ silicon powder with size distribution $<30$ $\mu$m (green).}}
  \end{figure}

Figure~\ref{fig:silica_trans_photos} shows transmission spectra and photographs of silica aerogel samples. The transmission spectra demonstrate high transmission at millimeter wavelengths, rejection of IR power at $> 25$ dB, and a tunable cutoff frequency by choice of silicon loading density and particle size distribution. Figure~\ref{fig:polyimide_vna_meas} summarizes measurements of polyimide aerogel samples with the VNA setup, demonstrating high transmission ($>99$\%) and low reflection ($\sim - 25$ dB or 0.3\%) at W band. Figure~\ref{fig:polyimide_trans} shows FTS transmission spectra in the IR across the range 1--15 THz for polyimide Samples 8, 9, and 10, while Figure~\ref{fig:polyimide_photo} shows a photograph of one of the polyimide filters mounted on a 40-cm ring for installation in a test receiver. 

\subsection{\added{Agreement with Models}}
\label{sec:model_meas}
\added{The integrated model of Sec~\ref{sec:integrated_model} is used primarily to give an initial guide in choosing aerogel base density, scattering particle size distribution, and scattering particle density to achieve a desired cutoff frequency and low-frequency effective index of refraction. From Figure~\ref{fig:mm_trans_model}, it is clear that there are some discrepancies between the model and measurements. Cutoff frequencies estimated from the model are typically only within $\pm 50$ \% of design values; however, the cutoff frequency between samples follows expected trends in which cutoff frequency increases with decreasing particle size, particle loading density, and/or filter thickness. Once a given formulation has been measured, this allows future formulations to be tuned to more precisely place the cutoff frequency at the desired value. }

\added{In addition, the cutoff is observed to be broader in measurements than predicted by the theory. Both the shift in measured cutoff frequency from the design value and the broader cutoff could be caused by a breakdown of the assumptions made in the model about either the distribution of particles sizes or the spherical shape of the particles in Mie scattering theory. The models plotted in Figure~\ref{fig:mm_trans_model} assume a uniform distribution of particle sizes within the range given, whereas in reality this distribution could be skewed or there could be undesired inclusion of smaller particles that did not get removed fully in the sieving process. Future work will attempt to measure particle properties in more detail before incorporation into filters to address these concerns. 
}

\added{Future work will more fully characterize filter optical performance, including measurement and modeling of diffuse transmission and scattering. A cryogenic far-IR integrating sphere is being commissioned for use with an existing FTS to measure diffuse transmission and scattering in these materials.}~\cite{Quijada2011} 
\added{This data can be used to separate absorption from diffuse scattering,}~\cite{Zhao2016, Lallich2009} \added{giving a more complete picture of the optical behavior and heat transport behavior of aerogel scattering filters above the cutoff frequency that includes the effect of haze.}~\cite{Zhao2019, Yu2014}

\subsection{\added{Cryogenic Operation}}
\added{Aerogel scattering filters have been tested for survivability at cryogenic temperatures. Samples of both silica and polyimide aerogel, with and without particle loading, have been cryogenically cycled to $\sim 70$~K between 5 and 10 times, depending on the sample, and have shown no degradation. Optical properties of filter samples have not been tested cryogenically; however, limited changes are expected in filter cutoff and transmission from the modest, $\lesssim 10^{-2}$, fractional change in dimensions, density, and index of refraction associated with thermal contraction. }

\section{\added{Discussion of Implementation and Limits of Operation}}
\added{We initially investigated silica aerogel as a base for scattering filters due to its wide availability and well-studied properties; however, challenges arose in using silica aerogel as a filter substrate material due to its brittleness and tendency to fracture. As mentioned in Section~\ref{sec:fabrication}, two silica aerogel densities were investigated, 50 and 90 mg/cm$^3$. The 50 mg/cm$^3$ samples were deemed too fragile to successfully use, as they tended to fracture when being mounted to measurement fixtures. The 90 mg/cm$^3$ samples were able to be successfully mounted and may be the preferred material system for filter apertures up to at least 10 cm, given the ability to achieve lower index of refraction than the polyimide aerogels investigated (See Table~\ref{tbl:samples}). For our specific application, the brittleness of silica aerogel was a concern, as we required filters with diameters of up to 40 cm that could be successfully used cryogenically. Mounting arrangements for large-format silica aerogel that would be compliant to differential contraction between the aerogel and metal mounting rings were deemed too risky. This lead to investigation of polyimide aerogel substrates, which are mechanically robust at 40~cm and flexible enough to be compliant to differential thermal contraction, while maintaining acceptable optical performance. The mechanical robustness of polyimide aerogels make them the preferred material system for a majority of applications, but other considerations, such as the possibility that polyimide aerogel would be etched by O$_2$ plasma in low-Earth orbit, might suggest use of silica or other aerogel formulations for specific applications.}~\cite{O2_etching_paper} 

\added{Given the small ($\sim 20$~nm typical) pore size of aerogels, the useful range of frequencies for aerogel scattering filters in principle extends into the visible. For frequencies at the blue end of the visible or into the ultraviolet, intrinsic scattering from the aerogel pore structure leads to haze.}~\cite{Zhao2019} \added{However, extending use of aerogel scattering filters to higher frequencies requires identification of aerogel formulations free from absorption bands at the frequencies of interest. Both silica and polyimide aerogels show strong absorption bands in the mid-IR above 10~THz, limiting their use to lower frequencies. }

\section{Conclusion}
We have fabricated infrared blocking filters composed of ultra-low index of refraction aerogel substrates with embedded silicon scattering particles. Both silica and polyimide aerogel filters were fabricated and tested for transmission and reflection at millimeter and far-IR wavelengths. We have demonstrated the ability of aerogel scattering filters to have (1) high ($> 99$\%) transmission across an ultra-broad bandwidth from DC to 1 THz or higher, (2) $> 25$ dB rejection of IR power, and (3) a cutoff frequency that is tunable through choice of silicon particle size and loading density. 

\section{Acknowledgements}
We acknowledge the National Science Foundation Division of Astronomical Sciences for their support of this work under Grant Numbers 0959349, 1429236, 1636634, and 1654494, as well as the National Aeronautics and Space Administration under grant number NNX14AB76A. We would like to thank Dipanjan Chaudhuri and others in the Armitage lab at Johns Hopkins University for making measurements of silica aerogel filter samples; Ocellus Inc. for providing silica aerogel samples; and Kevin Miller and Alyssa Barlis in the Optics Branch at GSFC for FTS measurements of polyimide aerogel samples. T. E.-H. was funded by a National Science Foundation Astronomy and Astrophysics Postdoctoral Fellowship. 

\section{Disclosures}
The authors declare no conflicts of interest to disclose. 

\newpage
\appendix
\section{Scattering Theory}\label{sec:scattering_theory}
As developed in Refs.~\citenum{1957lssp.book.....V, 1983asls.book.....B}, for spherical particles $Q_{\text{sca}}$ is given by 

\begin{equation}
Q_{\text{sca}} = \frac{2}{x^{2}} \sum^{\infty}_{n=1} \left( 2 n + 1 \right) \left( \left| a_{n} \right|^{2} + \left| b_{n} \right|^{2} \right) ,
\end{equation}

\noindent where $x=2\pi a /\lambda$ is the size parameter that depends on the radius of the particle, $a$, and the wavelength of the light, $\lambda$. The coefficients $a_{n}$ and $b_{n}$ are given in terms of Riccati-Bessel functions as

\begin{equation}
a_{n} = \frac{\varPsi^{\prime}_{n} (m x)\varPsi_{n} (x) - m \varPsi_{n} (m x)\varPsi^{\prime}_{n} (x)}{\varPsi^{\prime}_{n} (m x)\zeta_{n} (x) - m \varPsi_{n} (m x)\zeta^{\prime}_{n} (x)}
\end{equation}

\noindent and

\begin{equation}
b_{n} = \frac{m \varPsi^{\prime}_{n} (m x)\varPsi_{n} (x) - \varPsi_{n} (m x)\varPsi^{\prime}_{n} (x)}{m \varPsi^{\prime}_{n} (m x)\zeta_{n} (x) - \varPsi_{n} (m x)\zeta^{\prime}_{n} (x)} .
\end{equation}

\noindent Primes denote derivatives with respective to $x$, and the functions $\varPsi_{n}$ and $\zeta_{n}$ are written in terms of Bessel functions of the first kind, $J_{n}$, and Hankel functions of the second kind, $H_{n}^{(2)}$,

\begin{equation}
\begin{array}{c}
\varPsi_{n} (x) = \left( \frac{\pi x}{2} \right)^{1/2} J_{n+1/2} (x) ; \\ \\ \zeta_{n} (x) = \left( \frac{\pi x}{2} \right)^{1/2} H^{(2)}_{n+1/2} (x) .
\end{array}
\end{equation}

\bibliography{si_powder}

\end{document}